	\documentclass[%
	reprint,
	superscriptaddress,
	showpacs,
	 amsmath,amssymb,
	 aps,
	 pre,
	]{revtex4-1}
	
	\usepackage{graphicx}
	\usepackage{subcaption}
	\usepackage{caption}
	\usepackage{dcolumn}
	\usepackage{bm}
	\usepackage[inline]{enumitem}
	\usepackage{multirow}
	
	\newcommand{\myeqref}[1]{Eq.~\eqref{#1}}
	\newcommand{\mySecRef}[1]{Sec.~\ref{#1}}
	
\begin{document}

	\title{On the derivations of lattice Boltzmann evolution equation}

	\author{Huanfeng Ye}
	\email [E-mail address: ]{bdqhswj@sjtu.edu.cn}
	\affiliation{School of Nuclear Science and Engineering, Shanghai Jiao Tong University, Shanghai 200240, China}
	
	\author{Bo Kuang}
	\email [E-mail address: ]{bkuang@sjtu.edu.cn}
	\affiliation{School of Nuclear Science and Engineering, Shanghai Jiao Tong University, Shanghai 200240, China}
	
	\author{Yanhua Yang}%
	\email [E-mail address: ]{yanhuay@sjtu.edu.cn}
	\affiliation{School of Nuclear Science and Engineering, Shanghai Jiao Tong University, Shanghai 200240, China}
	\affiliation{National Energy Key Laboratory of Nuclear Power Software, Beijing 102209, China}

	\date{\today}
	
	\begin{abstract}
	A comparative analysis on the popular schemes for evaluating evolution equation in lattice Boltzmann method (LBM) is presented in this paper. It includes two classical characteristic-line schemes, Boesh-Karlin and He-Luo scheme, and a author-proposed scheme, Taylor-expansion scheme, originating from the extension of He-Luo scheme. We detailly discuss the mathematical mechanism and the equilibrium distribution evolution behind them. By analyzing the conflict between prediction and derivation, we address the preconditions for these schemes. At the end, we conclude their pros and cons and suggest  scheme's applicable scene based on their derivation procedure and further development capacity.

	\end{abstract}
	
	
	\maketitle
	
	
	\section{Introduction}
	\label{sec:intro}
	
	In the past decades, numerous works based on lattice Boltzmann method (LBM) have been published, involving almost all areas in computational fluid dynamic (CFD),  such as hydrodynamic systems \cite{LiuZou-178,BaochangZhaoli-179}, multiphase and multicomponent fluids \cite{LaddVerberg-186,SheikholeslamiGorji-Bandpy-163,LiuCheng-181,SemmaElGanaoui-240,MountrakisLorenz-483}, porous media flow \cite{MengGuo-2046,LiuHe-2047}.
	They all benefit from the development of LBM theories, especially the theories based on BGK-Boltzmann equation \cite{BhatnagarGross-95}. 
	LBM is considered as a mesoscopic algorithm, which calculates a distribution involves both micro and macro physics. 
	The theory of LBM consists with three parts: 
	\begin{enumerate*}[label={\bfseries \Roman*. }]
		\item evaluating evolution equation from kinetic theory \cite{BoeschKarlin-2075,HeChen-173,HeLuo-2016,YeKuang-2092};
		\item discretizing equilibrium distribution \cite{Shan-2088,HeLuo-2016,YeKuang-2092};
		\item recovering macroscopic equations \cite{Hwang-2023,YongZhao-2019,HeChen-173,YeKuang-2092}.
	\end{enumerate*}
	The evaluation of  evolution equation from kinetic theory illustrates its microphysical process. It attracts lots of scholars, referring to the aforementioned cites. A major branch is integrating the characteristic-line BGK-Boltzmann equation.

	In this paper, we intend to take a comparative analysis on the popular characteristic-line schemes. 
	The analysis tends to offer a profound understanding of the mechanism behind the schemes and discuss the their applicable scene.
	We will analyze three schemes, including two popular schemes under characteristic line LBM theory, Boesch-Karlin (BK) \cite{BoeschKarlin-2075} and He-Luo (HL) \cite{HeLuo-2016} scheme, and a author-proposed scheme, Taylor-expansion (TE) scheme, originating from the extension of BK scheme. 
	These three schemes can be classified into two kinds: partial integration approach and linear first-order ordinary differential equation approach.  Our argument will show that both kinds try to analytically integrate BGK-Boltzmann equation along the characteristic line. 
	The distinction is that they employ different expressions of the equilibrium distribution evolution along the characteristic line. 
	We will analyze the cause of the conflict on schemes' evolution equations, which leads us to the preconditions of schemes.
	Then we  will comment on schemes' derivation procedure and their further development capacity, and suggest their applicable scene.

	The remainder of this paper is organized as follows. \mySecRef{sec:bfintr} offers a brief review of the characteristic-line LBM theory, focusing on the evaluation of evolution equation. \mySecRef{sec:BKLB} provides a detail derivation of the BK scheme. We revisit the procedure of BK scheme and demonstrate the mathematical mechanism behind it. \mySecRef{sec:TELB} proposes TE scheme as an extension of the mathematical skill, partial integration, behind BK scheme. The derivation employs the same condition with BK scheme. \mySecRef{sec:HLLB} reviews the derivation of HL scheme and recovers all the evolution equations in BK and TE scheme. \mySecRef{sec:AppCon} presents an analysis of the conflict on the derived evolution equations. Based on the analysis, we address the preconditions for schemes. \mySecRef{sec:cpAna} concludes the pros and cons of schemes. The conclusion is argued in two way: the performance in previous derivation and further development capacity. Finally, \mySecRef{sec:CON} summarizes the whole paper.

	\section{Brief review on characteristic-line LBM theory}
	\label{sec:bfintr}
	Before we start the detail derivation and analysis, we would like to give a brief review of characteristic-line LBM theory \cite{YeKuang-2092,Dellar-2074,HeLuo-2016}.
	There are two keynotes in the characteristic-line LBM theory:
	\begin{enumerate*}[label={\bfseries \Roman*. }]
		\item employing characteristic-line integral to evaluate the evolution equation from BGK-Boltzmann equation, which establishes the relationship between LBM and classical kinetic theory;
		\item recovering the macroscopic equations directly from BGK-Boltzmann equation regardless of the form of evolution equations, which constructs the relationship between LBM and real physical world.
	\end{enumerate*}
	In this section, the review will focus on the evaluation of evolution equation, which identifies a scheme.
	
	We start with the BGK-Boltzmann equation \cite{BhatnagarGross-95}.
	\begin{equation}\label{eq:bgk}
	\frac{{\partial f}}{{\partial t}} + \vec \xi  \cdot \frac{{\partial f}}{{\partial \vec r}} =  - \frac{1}{\lambda }\left( {f - g} \right)
	\end{equation}
	where 
	$f \equiv f\left( {\vec r,\vec \xi ,t} \right)$ 
	is the particle distribution, 
	$\vec r$ and $t$ are the location and time respectively,
	$\vec \xi $ 
	is particle's microscopic velocity, 
	$\lambda $ 
	is the collision time, and $g \equiv g\left( {\vec r,\vec \xi ,t} \right) $
	is the Maxwell-Boltzmann distribution.
	\begin{equation}\label{eq:maxw}
	g \equiv \frac{\rho }{{{{\left( {2\pi RT} \right)}^{D/2}}}}\exp \left( { - \frac{{{{\left( {\vec \xi  - \vec u} \right)}^2}}}{{2RT}}} \right)
	\end{equation}
	where $R$ is the ideal gas constant, $D$ is the dimension of the space, and $\rho $, $\vec u$, and $T$ are the macroscopic density of mass, velocity and temperature, respectively.
	
	Along the characteristic line $\vec{r}+\vec{\xi} t$, the partial differential equation, BGK-Boltzmann equation, becomes an ordinary differential equation (ODE), in which
	the left-hand side in \myeqref{eq:bgk} can be interpreted as the total differential of $f$ with respect to $t$,
	\begin{equation}\label{eq:CLbgk}
	\frac{{df}}{{dt}} =  - \frac{1}{\lambda }\left( {f - g} \right)
	\end{equation}
	where
	\[\frac{d}{{dt}} \equiv \frac{\partial }{{\partial t}} + \vec \xi  \cdot \frac{\partial }{{\partial \vec r}}\]
	Integrating \myeqref{eq:CLbgk} along the characteristic-line $\vec{r}+\vec{\xi}t$ with respect to $t$, we can reach
	\begin{align}
	\label{eq:MCLBM}
	&f\left( {\vec r + \Delta t\vec \xi ,\vec \xi ,t + \Delta t} \right) - f\left( {\vec r,\vec \xi ,t} \right) = \nonumber \\
	&   - \frac{1}{\lambda }\int\limits_0^{\Delta t} {\left( {f\left( {\vec r + \vec \xi t',\vec \xi ,t + t'} \right) - g\left( {\vec r + \vec \xi t',\vec \xi ,t + t'} \right)} \right)dt'} 
	\end{align}
	In order to make notations lighter, short-hand expressions will be hired. Furthermore, we denote
	\begin{align*}
	{f^{n + 1}} &= f\left( {\vec r + \Delta t\vec \xi ,\vec \xi ,t + \Delta t} \right)\\
	{f^n} &= f\left( {\vec r,\vec \xi ,t} \right)\\
	{g^{n + 1}} &= g\left( {\vec r + \Delta t\vec \xi ,\vec \xi ,t + \Delta t} \right)\\
	{g^n} &= g\left( {\vec r,\vec \xi ,t} \right)\\
	f\left( {t'} \right) &= f\left( {\vec r + \vec \xi t',\vec \xi ,t + t'} \right)\\
	g\left( {t'} \right) &= g\left( {\vec r + \vec \xi t',\vec \xi ,t + t'} \right)
	\end{align*}
	Thus, the integral along the characteristics, \myeqref{eq:MCLBM},  can be rewritten as
	\begin{equation}
	\label{eq:SMCLBM}
	{f^{n + 1}} - {f^n} =  - \frac{1}{\lambda }\int\limits_0^{\Delta t} {\left( {f\left( {t'} \right) - g\left( {t'} \right)} \right)dt'} 
	\end{equation}
	Employing assumptions and integral
	 techniques on the right-hand side of \myeqref{eq:SMCLBM},  proposed by schemes \cite{BoeschKarlin-2075,HeChen-173,HeLuo-2016,YeKuang-2092}, the corresponding LBM evolution equation will be integrated.

	\section{Boesch-Karlin scheme}
	\label{sec:BKLB}
	Boesch-Karlin (BK) scheme employs Euler-Maclaurin integral to integrate the collision term along the characteristic line \cite{BoeschKarlin-2075}. 
	The scheme is  quite elegant and profound. 
	It not only offers an analytical solution for the equation \myeqref{eq:SMCLBM}, but also recovers the scheme proposed by He, Chen and Doolen (HCD) \cite{HeChen-173} and estimates its residual error, which is another milestone in the development of characteristic-line LBM theory.
	In this section, we will revisit their approach. For the sake of further discussion, the detail will be a bit different. The integral in this section would apply on the left-hand side of the BGK-Boltzmann equation instead of the collision term in original BK scheme.
	
	Due to applicable interval of Euler-Maclaurin Integral \cite{ArfkenWeber-2095}, the BGK-Boltzmann equation  would firstly be parametrized with a non-dimensional $s \in \left[ {0,1} \right]$
	\begin{equation}
	\label{eq:pBGK}
	\frac{d}{{ds}}\chi\left( s \right)  =  - \frac{{\Delta t}}{\lambda }\left( {\chi\left( s \right)  - \varphi\left( s \right) } \right)
	\end{equation} 
	where
	\[\begin{array}{l}
	\chi \left( s \right) = f\left( {\Delta ts } \right)\\
	\varphi \left( s \right) = {g}\left( {\Delta ts } \right)
	\end{array}\]
	To simplify the notation, we denote the subtraction term as $Q$ in the following paper,
	\begin{equation}
	Q\left( s \right) = \chi \left( s \right) - \varphi \left( s \right)
	\end{equation}
	
	Now integrating the left-hand side in \myeqref{eq:pBGK} with respect to $s$ from $0$ to $1$, instead of directly integrating, we employ the subsection integral by inducing Bernoulli polynomials as an artificial integrating part.
	\begin{align}
	\label{eq:SIBGK}
	\int\limits_0^1 {\frac{{d\chi \left( s \right)}}{{ds}}ds}  &= \int\limits_0^1 {\frac{{d\chi \left( s \right)}}{{ds}}{B_0}\left( s \right)ds}  \nonumber \\
	&= \left. {\frac{{{B_1}\left( s \right)}}{1}\frac{{d\chi \left( s \right)}}{{ds}}} \right|_0^1 - \int\limits_0^1 {\frac{{{d^2}\chi \left( s \right)}}{{d{s^2}}}\frac{{{B_1}\left( s \right)}}{1}ds}
	\end{align} 
	With the property of Bernoulli polynomials,
	\begin{align*}
		&{B_{n - 1}}\left( s \right) = \frac{1}{n}\frac{{d{B_n}\left( s \right)}}{{ds}} \\
		&{B_n}\left( 1 \right) = {\left( { - 1} \right)^n}{B_n}\left( 0 \right) = {\left( { - 1} \right)^n}{b_n}\\
		&{b_{2n + 1,n > 0}} = 0
	\end{align*}
	where $b_n$ is the Bernoulli number, we repeat the subsection integral on \myeqref{eq:SIBGK}, which leads us to Euler-Maclaurin integration.
	\begin{align}
		\label{eq:EMExp}
		\int\limits_0^1 {\frac{{d\chi \left( s \right)}}{{ds}}ds}  &= \sum\limits_{n = 1}^\infty  {\frac{{{{\left( { - 1} \right)}^{n - 1}}{b_n}}}{{n!}}\left( {{{\left( { - 1} \right)}^n}{{\frac{{{d^n}\chi }}{{d{s^n}}}}_{s = 1}} - {{\frac{{{d^n}\chi }}{{d{s^n}}}}_{s = 0}}} \right)} \nonumber \\
		& =  - {b_1}\left( {{{\frac{{d\chi }}{{ds}}}_{s = 1}} + {{\frac{{d\chi }}{{ds}}}_{s = 0}}} \right) \nonumber \\ 
		& - \sum\limits_{n = 1}^\infty  {\frac{{{b_{2n}}}}{{\left( {2n} \right)!}}\left( {{{\frac{{{d^{2n}}\chi }}{{d{s^{2n}}}}}_{s = 1}} - {{\frac{{{d^{2n}}\chi }}{{d{s^{2n}}}}}_{s = 0}}} \right)}
	\end{align}
	It should be noted that until now, the derivation is independent from BGK collision model. Applying the relation in \myeqref{eq:pBGK}, the $n$th derivative of $\chi$ can be expressed as,
	\begin{equation}
	\label{eq:EMterm}
	\frac{{{d^n}\chi \left( s \right)}}{{d{s^n}}} = {\left( { - \frac{{\Delta t}}{\lambda }} \right)^n}Q\left( s \right) - \sum\limits_{i = 1}^{n - 1} {{{\left( { - \frac{{\Delta t}}{\lambda }} \right)}^{n - i}}\frac{{{d^i}\varphi \left( s \right)}}{{d{s^i}}}}
	\end{equation}
	The above formula contains a tricky term, the derivatives of $\varphi$ with respect to $s$, called as the propagation-type term by BK scheme. By estimating the magnitude of the propagation-type term, which is the derivatives of the locally conserved fields along the characteristics, BK scheme ignored it during processing the LBM evolution equation. This dealing can be interpreted as an assumption,
	\begin{equation}
	\label{eq:EMAssp}
	{\frac{{{d^i}\varphi \left( s \right)}}{{d{s^i}}}_{i > 0}} = 0
	\end{equation}
	Then \myeqref{eq:EMExp} can be rewritten as
	\begin{align}
		\label{eq:EMLB}
		\int\limits_0^1 {\frac{{d\chi \left( s \right)}}{{ds}}ds} & = {b_1}\left( {\frac{{\Delta t}}{\lambda }} \right)\left( {Q\left( 1 \right) + Q\left( 0 \right)} \right)\nonumber \\
		& - \sum\limits_{n = 1}^\infty  {\frac{{{b_{2n}}}}{{\left( {2n} \right)!}}{{\left( { - \frac{{\Delta t}}{\lambda }} \right)}^{2n}}\left( {Q\left( 1 \right) - Q\left( 0 \right)} \right)} 	
	\end{align}
	Inducing the generating function of Bernoulli number $b_n$ \cite{ArfkenWeber-2095},
	\begin{equation}
	\frac{t}{{{e^t} - 1}} = \sum\limits_{n = 1}^\infty  {\frac{{{b_{2n}}{t^{2n}}}}{{\left( {2n} \right)!}} + {b_0} + {b_1}t} 
	\end{equation}
	and implementing the direct integral on the left-hand side of \myeqref{eq:EMLB}, the equation turns into
	\begin{align}
	\chi \left( 1 \right){\rm{ }} - \chi \left( 0 \right) = & - \frac{1}{2}\left( {\frac{{\Delta t}}{\lambda }} \right)\left( {Q\left( 1 \right) + Q\left( 0 \right)} \right) \nonumber \\
	& - \left( {\frac{{ - \frac{{\Delta t}}{\lambda }}}{{{e^{ - \frac{{\Delta t}}{\lambda }}} - 1}} - 1 - \frac{{\Delta t}}{{2\lambda }}} \right)\left( {Q\left( 1 \right) - Q\left( 0 \right)} \right)
	\end{align}
	Using the non-parametrized notation, the integration of BGK-Boltzmann equation with Euler-Maclaurin integral, or the exact lattice Boltzmann(LB) equation called by BK scheme, is recovered.
	\begin{align}
		\label{eq:EMLBM}
	&{f^{n + 1}} - {f^n} =  - \frac{1}{2}\left( {\frac{{\Delta t}}{\lambda }} \right)\left( {\left( {{f^{n + 1}} - {g^{n + 1}}} \right) + \left( {{f^n} - {g^n}} \right)} \right) \nonumber \\
	&- \left( {\frac{{ - \frac{{\Delta t}}{\lambda }}}{{{e^{ - \frac{{\Delta t}}{\lambda }}} - 1}} - 1 - \frac{{\Delta t}}{{2\lambda }}} \right)\left( {\left( {{f^{n + 1}} - {g^{n + 1}}} \right) - \left( {{f^n} - {g^n}} \right)} \right)
	\end{align}
	When we only take the first term in \myeqref{eq:EMLBM} as the collision integral, the equation turns into the HCD equation. And its residual error is the second term of the right-hand side in \myeqref{eq:EMExp}. Reconstructing the \myeqref{eq:EMLBM} in the form of evolution equation, we can obtain
	\begin{align}
		\label{eq:BKEQ}
	{f^{n + 1}} - {f^n} &= -\left( {1 - {e^{ - \frac{{\Delta t}}{\lambda }}} } \right)\left( {{f^n} - {g^n}} \right) \nonumber \\
	&+ \left( {1 + \frac{\lambda }{{\Delta t}}\left( {{e^{ - \frac{{\Delta t}}{\lambda }}} - 1} \right)} \right)\left( {{g^{n + 1}} - {g^n}} \right)
	\end{align}  
	
	\section{Taylor-expansion scheme}
	\label{sec:TELB}
	In \mySecRef{sec:BKLB}, we have illustrated the methodology of evaluating LBM evolution equation with partial integration. It should be noted that the Bernoulli polynomials is merely one of the available choices for the artificial integrating part. The design of the artificial integration part is quite arbitrary. The only requirement is 
	\begin{align}
		{P_0}\left( t \right) &= 1 \nonumber \\
		{P_{n - 1}}\left( t \right) &= c_n\frac{{d{P_n}\left( t \right)}}{{dt}}
	\end{align}
	where $c_n$ is a coefficient.Here we offer another two series of polynomials,$P_n^1\left( t \right)$ and $P_n^2\left( t \right)$, 
	\begin{subequations}
		\begin{align}
			P_n^1\left( t \right) &= {t^n}\\
			P_n^2\left( t \right) &= {\left( {t - \Delta t} \right)^n}
		\end{align}
	\end{subequations}
	These two polynomial series can be directly applied on the original BGK-Boltzmann equation, \myeqref{eq:CLbgk}, without parametrization. Repeating the procedure demonstrated in \mySecRef{sec:BKLB}, the corresponding LBM evolution equations can be easily resolved.
	\begin{subequations}
		\label{eq:pLBM}
		\begin{align}
			{f^{n + 1}} - {f^n} &= \sum\limits_{n = 1}^\infty  {\left. {{{\left( { - 1} \right)}^{n - 1}}\frac{1}{{n!}}\frac{{{d^n}f\left( {t'} \right)}}{{d{t^n}}}P_n^1\left( {t'} \right)} \right|} _{t' = 0}^{\Delta t}\nonumber \\
			&=  - \sum\limits_{n = 1}^\infty  {\frac{{{{\left( { - \Delta t} \right)}^n}}}{{n!}}\frac{{{d^n}{f^{n + 1}}}}{{d{t^n}}}} \nonumber \\
			&=  - \left( {{e^{\Delta t/\lambda }} - 1} \right)\left( {{f^{n + 1}} - {g^{n + 1}}} \right) \label{eq:pLBMA}\\
			{f^{n + 1}} - {f^n} &= \sum\limits_{n = 1}^\infty  {\left. {{{\left( { - 1} \right)}^{n - 1}}\frac{1}{{n!}}\frac{{{d^n}f\left( {t'} \right)}}{{d{t^n}}}P_n^2\left( {t'} \right)} \right|} _{t' = 0}^{\Delta t}\nonumber \\
			&= \sum\limits_{n = 1}^\infty  {\frac{{{{\left( {\Delta t} \right)}^n}}}{{n!}}\frac{{{d^n}{f^n}}}{{d{t^n}}}} \nonumber \\
			&= \left( {{e^{ - \Delta t/\lambda }} - 1} \right)\left( {{f^n} - {g^n}} \right)\label{eq:pLBMB}
		\end{align}
	\end{subequations}
	where
	\begin{align*}
	\frac{{{d^n}{f^{n + 1}}}}{{d{t^n}}}& = {\left( { - \frac{1}{\lambda }} \right)^n}\left( {{f^{n + 1}} - {g^{n + 1}}} \right) \nonumber \\
	&\qquad- \sum\limits_{i = 1}^{n - 1} {{{\left( { - \frac{1}{\lambda }} \right)}^{n - i}}\frac{{{d^i}{g^{n + 1}}}}{{d{t^i}}}} \nonumber \\
	&\cong {\left( { - \frac{1}{\lambda }} \right)^n}\left( {{f^{n + 1}} - {g^{n + 1}}} \right) \nonumber \\
	\frac{{{d^n}{f^n}}}{{d{t^n}}}& = {\left( { - \frac{1}{\lambda }} \right)^n}\left( {{f^n} - {g^n}} \right) \nonumber \\
	&\qquad - \sum\limits_{i = 1}^{n - 1} {{{\left( { - \frac{1}{\lambda }} \right)}^{n - i}}\frac{{{d^i}{g^n}}}{{d{t^i}}}} \nonumber \\
	& \cong {\left( { - \frac{1}{\lambda }} \right)^n}\left( {{f^n} - {g^n}} \right)
	\end{align*}

	As the partial integration with Bernoulli polynomials generates the Euler-Maclaurin integral form of BGK-Boltzmann equation, the above polynomial series in fact generate the Taylor expansions (TE) on different reference points. The prove is quite simple. Expanding $f^n$ and $f^{n+1}$ at reference points $f^{n+1}$ and $f^n$ respectively, they become
	\begin{subequations}
		\label{eq:TE}
		\begin{align}
			{f^n} &= \sum\limits_{n = 0}^\infty  {\frac{{{{\left( { - \Delta t} \right)}^n}}}{{n!}}\frac{{{d^n}{f^{n + 1}}}}{{d{t^n}}}}  \label{eq:TEA}\\
			{f^{n + 1}} &= \sum\limits_{n = 0}^\infty  {\frac{{{{\left( {\Delta t} \right)}^n}}}{{n!}}\frac{{{d^n}{f^n}}}{{d{t^n}}}} \label{eq:TEB}
		\end{align}
	\end{subequations}
	Substituting the expansion into the left-hand side of \myeqref{eq:pLBM} correspondingly,  then the derived equations are directly recovered.
	\begin{subequations}
		\label{eq:TEPR}
		\begin{align}
			{f^{n + 1}} - {f^n} &= {f^{n + 1}} - \sum\limits_{n = 0}^\infty  {\frac{{{{\left( { - \Delta t} \right)}^n}}}{{n!}}\frac{{{d^n}{f^{n + 1}}}}{{d{t^n}}}} \nonumber \\
			&=  - \sum\limits_{n = 1}^\infty  {\frac{{{{\left( { - \Delta t} \right)}^n}}}{{n!}}\frac{{{d^n}{f^{n + 1}}}}{{d{t^n}}}}   \label{eq:TEPRA}\\
			{f^{n + 1}} - {f^n} &= \sum\limits_{n = 0}^\infty  {\frac{{{{\left( {\Delta t} \right)}^n}}}{{n!}}\frac{{{d^n}{f^n}}}{{d{t^n}}}}  - {f^n} \nonumber \\ 
			&= \sum\limits_{n = 1}^\infty  {\frac{{{{\left( {\Delta t} \right)}^n}}}{{n!}}\frac{{{d^n}{f^n}}}{{d{t^n}}}} \label{eq:TEPRB}
		\end{align}
	\end{subequations}
	To distinguish the formulas in \myeqref{eq:pLBM}, we refer to \myeqref{eq:pLBMA} and \myeqref{eq:pLBMB} as right-point and left-point TE scheme respectively, based on their reference points. 
	
	 \section{He-Luo scheme}
	 \label{sec:HLLB}
	Despite of the complicated expansion and transformation  BK and TE scheme has implemented, 
	both their targets are to evaluate the change of $f$ along the characteristic line, ${f^{n + 1}} - {f^n}$. 
	In BK and TE scheme, the change is calculated as an entity, the integration of  collision term along the characteristic line. It also can be resolved by calculating the value of $f^{n + 1}$, which is adopted by He-Luo (HL) scheme.
	Once we knew the value of $f^{n + 1}$,  we can easily evaluate the change by formula, ${f^{n + 1}} - {f^n}$. 
	To resolve $f^{n + 1}$, HL scheme employs a 
	technique for solving linear first-order ordinary differential equation (ODE). By introducing a integrating factor $e^{t/\lambda}$, the BGK-Boltzmann equation can be transformed into the exact form \cite{ArfkenWeber-2095}. 
	\begin{equation}
	\label{eq:SFBGK}
	\frac{{d\left( {{e^{t/\lambda }}f} \right)}}{{dt}} = \frac{1}{\lambda }{e^{t/\lambda }}{g}
	\end{equation}
	It should be noted that the integrating factor $e^{t/\lambda}$ is specially designed for BGK-Boltzmann equation. Once the collision model changed, this approach may invalidate, which means we can't find  an integrating factor to make equation exact. 
	Now if $g$ was knew and integrable along the characteristic line with respect to $t$, \myeqref{eq:SFBGK} can be directly integrated. With simple reconstruction on the integration, we will solve $f^{n + 1}$.
	
	In Ref.~\cite{HeLuo-2016}, He and Luo proposed a design of $g$ along the characteristic line,
	\begin{equation}
	\label{eq:BKAssp}
	g\left( {t'} \right) = {g^n} + \frac{{t'}}{{\Delta t}}\left( {{g^{n + 1}} - {g^n}} \right)
	\end{equation} 
	Then the corresponding integration of \myeqref{eq:SFBGK}  yields
	\begin{align}
	{e^{\Delta t/\lambda }}{f^{n + 1}} &- {f^n} = \left( {{e^{\Delta t/\lambda }} - 1} \right){g^n} \nonumber \\
	&+ \left( {{e^{\Delta t/\lambda }} - \frac{\lambda }{{\Delta t}}{e^{\Delta t/\lambda }} + \frac{\lambda }{{\Delta t}}} \right)\left( {{g^{n + 1}} - {g^n}} \right)	
		\end{align}
	Transforming it into the form of evolution equation, we can obtain
	\begin{align}
		\label{eq:RECBK}
	{f^{n + 1}} - {f^n} = & - \left( {1 - {e^{ - \Delta t/\lambda }}} \right)\left( {{f^n} - {g^n}} \right)\nonumber \\
	&+ \left( {1 + \frac{\lambda }{{\Delta t}}\left( {{e^{ - \Delta t/\lambda }} - 1} \right)} \right)\left( {{g^{n + 1}} - {g^n}} \right)
	\end{align}
	which is exactly the same with the evolution equation of BK scheme, \myeqref{eq:BKEQ}. 
	
	What's more, from the forms of TE evolution equations, \myeqref{eq:pLBM}, we can easily recover the corresponding evolution function of $g\left( {t'} \right)$ behind them
	\begin{equation}
	\label{eq:TEAsp}
	g\left( {t'} \right) = \left\{ {\begin{array}{*{20}{l}}
		{{{g^{n + 1}}}}&{\mbox{For \myeqref{eq:pLBMA}}}\\
		{{{g^n}}}&{\mbox{For \myeqref{eq:pLBMB}}}
		\end{array}} \right.
	\end{equation}
	Integrating \myeqref{eq:SFBGK} with the assumption \myeqref{eq:TEAsp}, we can get
	\begin{subequations}
		\label{eq:TEIn}
	\begin{align}
	{e^{\Delta t/\lambda }}{f^{n + 1}} - {f^n} &= \left( {{e^{\Delta t/\lambda }} - 1} \right){g^{n + 1}}\\
	{e^{\Delta t/\lambda }}{f^{n + 1}} - {f^n} &= \left( {{e^{\Delta t/\lambda }} - 1} \right){g^n}
	\end{align}
		\end{subequations}
	Reconstructing \myeqref{eq:TEIn} in the form of evolution equation, they yield 
	\begin{subequations}
	\begin{align}
	{f^{n + 1}} - {f^n} &=  - \left( {{e^{\Delta t/\lambda }} - 1} \right)\left( {{f^{n + 1}} - {g^{n + 1}}} \right)\\
	{f^{n + 1}} - {f^n} &=  - \left( {1 - {e^{ - \Delta t/\lambda }}} \right)\left( {{f^n} - {g^n}} \right)
		\end{align}
	\end{subequations}
	They equate with the evolution equations of TE scheme, \myeqref{eq:pLBM}.

	\section{Preconditions for schemes}
	\label{sec:AppCon}
	In \mySecRef{sec:BKLB}, \mySecRef{sec:TELB} and \mySecRef{sec:HLLB}, we have illustrated three schemes to derive the evolution equation under characteristic-line LBM theory. The BK scheme and TE scheme are equivalent mathematically, which are applications of partial integration with different polynomial series. HL scheme is a bit different. It employs the strategy of solving the linear first-order ordinary differential equation. Comparing with BK and TE scheme, HL scheme is specifically designed for BGK-Boltzmann equation. All three schemes try to integrate the BGK-Boltzmann equation along the characteristic line analytically.  Under BGK collision model with the same $g\left( {t'} \right)$, all three schemes will converge on the same evolution equation, especially BK and TE scheme.
	
	Against our expectation,  our derivation shows that the mathematically equivalent BK and TE scheme diverge on evolution equations under the same condition.
	\begin{align}
		\label{eq:PIAssp}
		&\frac{{df}}{{dt}} =  - \frac{1}{\lambda }\left( {f - g} \right)\nonumber \\
		&{\frac{{{d^i}g}}{{d{t^i}}}_{i > 0}} = 0
	\end{align} 
	Even  employing the same TE scheme, the evolution equation varies with the reference point. 
	Luckily, we recover all evolution equations derived by BK and TE scheme with HL scheme. 
	Combining with the the HL scheme, we can analyze the cause of the conflict.

	We start with the formulas of TE scheme, \myeqref{eq:pLBM}. From \myeqref{eq:TEAsp} in HL scheme, we find that the same condition \myeqref{eq:PIAssp} has different functions for equilibrium distribution evolution $g\left( {t'} \right)$  under right-point and left-point TE scheme,
	\begin{subequations}
	\begin{align}
	g\left( {t'} \right) &= \left\{ {\begin{array}{*{20}{l}}
		{{g^{n + 1}},}&{\Delta t \ge t' > 0}  \\
		{{g^n},}&{t' = 0}
		\end{array}} \right.{\rm{ for ~ \myeqref{eq:pLBMA}}} \\
	g\left( {t'} \right)& = \left\{ {\begin{array}{*{20}{l}}
		{{g^{n + 1}},}&{t' = \Delta t} \\
		{{g^n},}&{\Delta t > t' \ge 0}
		\end{array}} \right.{\rm{ for ~ \myeqref{eq:pLBMB}}} 	
	\end{align}	
	\end{subequations}
	The difference arises from the discontinuity of \myeqref{eq:PIAssp}. The assumption of $g\left( {t'} \right)$ in \myeqref{eq:PIAssp} equates with
	\begin{align}
		\label{eq:PISF}
	&g\left( {t'} \right) = \left\{ {\begin{array}{*{20}{l}}
		{{g^{n + 1}},}&{t' = \Delta t}\\
		{{g^n},}&{t' = 0}
		\end{array}} \right. \nonumber \\
	&\frac{{dg\left( {t'} \right)}}{{dt'}} = 0
	\end{align}
	in which $g\left( {t'} \right)$ is a discontinuous step function. This discontinuity violates the application condition of Taylor expression. Once we reformed \myeqref{eq:PISF} in a continuous form,
	\begin{align}
		\label{eq:TECA}
		&g\left( {t'} \right) = c \nonumber \\
		&\frac{{dg\left( {t'} \right)}}{{dt'}} = 0
		\end{align}
	where $c$ is a constant, the right-point and left-point TE scheme generate the same evolution equation,
	\begin{equation}
	\label{eq:CATELBM}
	{f^{n + 1}} - {f^n} =  - \left( {1 - {e^{ - \Delta t/\lambda }}} \right)\left( {{f^n} - c} \right)
	\end{equation}
	This discontinuity also can explain the difference between TE and BK scheme. Introducing the assumption \myeqref{eq:TECA} into BK scheme, we can obtain
	\begin{align}
		\label{eq:CABKLBM}
	{f^{n + 1}} - {f^n} = & - \frac{1}{2}\left( {\frac{{\Delta t}}{\lambda }} \right)\left( {\left( {{f^{n + 1}} - c} \right) + \left( {{f^n} - c} \right)} \right) \nonumber \\
	& - \left( {\frac{{ - \frac{{\Delta t}}{\lambda }}}{{{e^{ - \frac{{\Delta t}}{\lambda }}} - 1}} - 1 - \frac{{\Delta t}}{{2\lambda }}} \right)\left( {{f^{n + 1}} - {f^n}} \right)
	\end{align}
	With transposition and simplification, the evolution equation \myeqref{eq:CATELBM} can be recovered from \myeqref{eq:CABKLBM}. It should be noted that the procedure of BK scheme is based on the parameterized BGK-Boltzmann equation, but for the convenience of discussion, we skip the parameterization and directly argue on original equilibrium distribution $g$. This convention would be kept in the following paper.
	
	Despite that BK scheme induced a discontinuous function of $g\left( {t'} \right)$ , the recovery of BK evolution equation under HL scheme shows it's not meaningless. The reason is the subtraction algorithm of high-order partial integral term in \myeqref{eq:EMExp},
	\begin{equation}
	\label{eq:EMHO}
	 - \sum\limits_{n = 1}^\infty  {\frac{{{b_{2n}}}}{{\left( {2n} \right)!}}\left( {{{\frac{{{d^{2n}}\chi }}{{d{s^{2n}}}}}_{s = 1}} - {{\frac{{{d^{2n}}\chi }}{{d{s^{2n}}}}}_{s = 0}}} \right)} 
	\end{equation}
	which eliminates the discontinuity implicitly. The subtraction form in \myeqref{eq:EMHO} makes the assumption, \myeqref{eq:EMAssp}, an over constraint. In fact, to derive the evolution equation of BK scheme, we only need
	\begin{equation}
	{\frac{{{d^i}g\left( {t'} \right)}}{{dt{'^i}}}_{t' = \Delta t}} = {\frac{{{d^i}g\left( {t'} \right)}}{{dt{'^i}}}_{t' = 0}},\quad {\rm{  for }}~i > 0
	\end{equation}
	Combining with the design of $g\left( {t'} \right)$, \myeqref{eq:BKAssp},  in derivation of \myeqref{eq:RECBK} in HL scheme, we can recover the implicit continuous assumption of $g\left( {t'} \right)$ behind BK scheme,
	 \begin{align}
	 	\label{eq:RBKAssp}
	 	&g\left( {t'} \right) = \left\{ {\begin{array}{*{20}{l}}
	 		{{g^{n + 1}}}&{t' = \Delta t}\\
	 		{{g^n}}&{t' = 0}
	 		\end{array}} \right. \nonumber \\
	 	&{\frac{{dg\left( {t'} \right)}}{{dt'}}_{t' = \Delta t}} = {\frac{{dg\left( {t'} \right)}}{{dt'}}_{t' = 0}}{\rm{  = }}\frac{1}{{\Delta t}}\left( {{g^{n + 1}} - {g^n}} \right) \nonumber \\
	 	&{\frac{{{d^i}g\left( {t'} \right)}}{{dt{'^i}}}_{t' = \Delta t}} = {\frac{{{d^i}g\left( {t'} \right)}}{{dt{'^i}}}_{t' = 0}} = 0, \quad {\rm{ for }}~i > 1
	 	\end{align} 
	In fact, \myeqref{eq:BKAssp} and \myeqref{eq:RBKAssp} describe the same evolution function of $g\left( {t'} \right)$ along the characteristic line with different expressions. Now we substitute the assumption \myeqref{eq:RBKAssp} into right-point and left-point TE scheme, \myeqref{eq:pLBM}, we can obtain
	\begin{subequations}
	\begin{align}
	{f^{n + 1}} - {f^n} = & - \left( {{e^{\Delta t/\lambda }} - 1} \right)\left( {{f^{n + 1}} - {g^{n + 1}}} \right) \nonumber \\
	 & + \frac{{ - \lambda }}{{\Delta t}}\left( {{e^{\Delta t/\lambda }} - 1 - \frac{{\Delta t}}{\lambda }} \right)\left( {{g^{n + 1}} - {g^n}} \right) \\
	{f^{n + 1}} - {f^n} = & \left( {{e^{ - \Delta t/\lambda }} - 1} \right)\left( {{f^n} - {g^n}} \right) \nonumber \\
	 & - \frac{{ - \lambda }}{{\Delta t}}\left( {{e^{ - \Delta t/\lambda }} - 1 + \frac{{\Delta t}}{\lambda }} \right)\left( {{g^{n + 1}} - {g^n}} \right)	
		\end{align}
	\end{subequations}
	With transposition and simplification,the above equations can be easily transformed into the evolution equation of BK scheme, \myeqref{eq:BKEQ}. It validates our statement that TE scheme is equivalent with BK scheme mathematically.

	Throughout our discussion on the difference among the evolution equations, the discontinuity of $g\left( {t'} \right)$ is the key reason. The derivation indicates that BK, TE and HL scheme solve the same equation with different expressions of $g\left( {t'} \right)$. For an instance, our derivation takes \myeqref{eq:BKAssp} under HL scheme while \myeqref{eq:RBKAssp} under BK and TE scheme. More precisely, in BK and TE scheme, the expression is a description of the property at the beginning and end points of characteristic line, while in HL scheme, it's the evolution function along the line. With invalid design of $g\left( {t'} \right)$, in which it's discontinuous under BK and TE scheme's expression, or non-integrable under HL scheme's, the evaluated evolution equation may be meaningless. Luckily, despite of employing the discontinuous assumption, the evolution equations of BK and TE scheme coincide with the solution of reasonable design, referring to the derivation in HL scheme. The only consequence is the divergence of evolution equations between BK and TE scheme. It's not fatal, but it warns us that there are preconditions for BK, TE and HL scheme. Only the prerequisites were satisfied, the evolution equation would be meaningful and explainable. Meanwhile,  BK, TE and HL scheme will converge on the same evolution equation, which could help to check our design.

	\section{comparative analysis on schemes}
	\label{sec:cpAna}
	As \mySecRef{sec:AppCon} has demonstrated, BK, TE and HL scheme are equivalent when they employ the same continuous assumption of $g\left( {t'} \right)$ along the characteristic line. 
	In BK and TE scheme's derivation, the procedure focuses on the property of $g\left( {t'} \right)$ at the beginning and end points instead of the evolution along the characteristic line, which tends to neglect the application conditions and induce inapplicable discontinuous $g\left( {t'} \right)$. 
	Comparing with TE scheme, BK scheme is a more special derivation, which would implicitly modify the introduced assumption due to the subtraction algorithm in high-order partial integral terms. For both BK and TE scheme, their description of $g\left( {t'} \right)$ makes it hard to image the physical procedure behind them.
	In contrast to BK and TE scheme, HL scheme pays more attention to the evolution form of $g\left( {t'} \right)$  along the characteristic line. It leads to a clear illustration of the evolution process along the characteristic line.
	For the same continuous $g\left( {t'} \right)$, the calculation of HL scheme is more concise than BK and TE scheme. In other word, HL scheme is better at dealing with model of $g\left( {t'} \right)$.
	
	It also should be noticed that the approach employed by BK and TE scheme is independent from BGK model while HL scheme is specially designed for it. The exact LB equation \myeqref{eq:EMLBM} is an analytical solution of BGK-Boltzmann equation with the assumption \myeqref{eq:RBKAssp} or \myeqref{eq:BKAssp}. Indeed, the "exact" denotes the analytically dealing with $f$ in the collision term along the characteristic line.
	Due to the analytically dealing with $f$, we expect the exact LB equation should have a better performance.
	 But the numerical result is quite against our wish. Inducing the trick proposed by Ref.~\cite{HeChen-173} to remove the implicitness, the exact LB equation yields 
	\begin{equation}
	\label{eq:exLBM}
	{h^{n + 1}} - {h^n} =  - \left( {1 - {e^{ - \Delta t/\lambda }}} \right)\left( {{h^n} - {g^n}} \right)
	\end{equation}
	where
	\begin{equation}
	\label{eq:bkExp}
	h = f + \left( {\frac{{\Delta t}}{\lambda }\frac{1}{{1 - {e^{ - \Delta t/\lambda }}}} - 1} \right)\left( {f - g} \right)
	\end{equation}
	Comparing with HCD equation,
	\begin{equation}
	\label{eq:HCDLBM}
	{h^{n + 1}} - {h^n} =  - \frac{1}{{0.5 + \lambda /\Delta t}}\left( {{h^n} - {g^n}} \right)
	\end{equation} 
	where
	\begin{equation}
	\label{eq:HCDExp}
	h = f + \frac{{\Delta t}}{{2\lambda }}\left( {f - g} \right)
	\end{equation}
	the exact LB equation is obviously unreliable. For details of the comparison, readers can refer to the numerical results of steady assumption (SA) model and evolutionary correction of derivation (ECD) model in Ref.~\cite{YeKuang-2092}. The calculations in Ref.~\cite{YeKuang-2092} is not based on the derivation of aforementioned schemes, but they employ the same corresponding LBM formulas, \myeqref{eq:exLBM} and \myeqref{eq:HCDLBM}, which ensures the validity of numerical results. What should be noted is that the comparison is based on the explicit distribution $h$, which ignores the difference of the definitions, \myeqref{eq:bkExp} and \myeqref{eq:HCDExp}. The effect of different expressions $h$ between exact LB and HCD equation, is a quite independent subject and beyond the scope of this paper. It should be addressed in a separate publication.

	The result of comparison reminds that BGK approximation is only a linear model of the original collision term in Boltzmann equation. The accuracy of LB equation depends on the calculation of whole collision term instead of $f$ in BGK model or even the BGK model itself. In this case, BK and TE scheme is more efficient than HL scheme. 
	In BK and TE scheme, the right-hand side of equation consists with derivatives of the collision term, independent from BGK model. And the terms lines up along the order of differentiation. It's convenient even straightforward to evaluate the collision integral by introducing new models or truncating the series, and estimate the residual error, referring to generating HCD equation from BK scheme \cite{BoeschKarlin-2075}. 
	On the contrary, HL scheme is bound with BGK model. For new collision models, it may encounter the challenge of applicability.  And the approximation of whole collision term with truncation would be a disaster, referring to the recovery of LBM equation in Ref.\cite{HeLuo-2016}

	We also would like to point out that BK and TE scheme converges at different rates in spite of their mathematical equivalence. BK scheme converges faster than TE scheme. It can be easily proven by their first-order approximation, for BK scheme, its residual error is 
	\begin{equation}
	\left| {\int\limits_0^1 {\frac{{d\chi }}{{ds}}ds}  - \frac{1}{2}\left( {{{\frac{{d\chi }}{{ds}}}_{s = 1}} + {{\frac{{d\chi }}{{ds}}}_{s = 0}}} \right)} \right| \le \frac{{{M_2}}}{{12}}
	\end{equation}
	where
	\[{M_2} = \mathop {\sup }\limits_{s \in [0,1]} \left| {\frac{{{d^2}}}{{d{s^2}}}\left( {\frac{{d\chi }}{{ds}}} \right)} \right|\]
	while for the left-point and right-point TE scheme, its residual error is
	\begin{subequations}
	\begin{align}
		\left| {\int\limits_0^{\Delta t} {\frac{{df}}{{dt}}dt}  - \Delta t{{\frac{{df}}{{dt}}}_{t = \Delta t}}} \right| \le \frac{{{M_1}}}{2} \\
		\left| {\int\limits_0^{\Delta t} {\frac{{df}}{{dt}}dt}  - \Delta t{{\frac{{df}}{{dt}}}_{t = 0}}} \right| \le \frac{{{M_1}}}{2} 
		\end{align}
	\end{subequations}
	where
	\[{M_1} = \mathop {\sup }\limits_{t \in [0,\Delta t]} \left| {\frac{d}{{dt}}\left( {\frac{{df}}{{dt}}} \right)} \right|\]
	For details, readers can refer to Riemann sum. 

In conclusion of this section, we have made a comment on the derivation procedure and further development capacity of schemes. Under BGK model, HL is a better solution to integrating the BGK-Boltzmann equation along the characteristic line. The calculation is straightforward and concise with a explicit physical picture.
Meanwhile the unreliable numerical performance of exact integration, the exact LB equation \myeqref{eq:exLBM}, reminds of the importance of overall collision term, which has been neglected in HL scheme. In this regard, BK and TE scheme are the better choice, which deal with the collision term as an entity and are independent from BGK model.
Furthermore, BK scheme converges faster than TE scheme.

	\section{conclusion}
	\label{sec:CON}

	In this paper, we have illustrated three schemes to derive the evolution equation along characteristic line. We started with the review of BK scheme. Employing the mathematical skill behind  BK scheme, partial integration, we proposed TE scheme and proved the equivalence between TE scheme and Taylor expansion. 
	Then we revisited HL scheme. In the demonstration of HL scheme, we gave all $g\left( {t'} \right)$ approximations behind BK and TE scheme. 
	By analyzing the cause of the conflict on evolution equations between BK and TE scheme, we argued the preconditions for BK, TE and HL scheme. Then we commented on the procedure and further development capacity of the schemes.

	Our investigation shows, under BGK model, BK, TE and HL scheme are equivalent, which are the analytical solution for integrating BGK-Boltzmann equation along the characteristic line. But there are preconditions for their implementation, the continuity or integrability of $g\left( {t'} \right)$. Once satisfying these prerequisites, BK, TE and HL will converge on the same evolution equation. 
	To our surprise, the derivation shows 
	that the BK evolution equation is a solution of linear equilibrium distribution evolution along the characteristic line. And its numerical performance is quite unreliable, which is against our original intention.
	Our demonstration also indicates that BK, TE and HL scheme have their own best applicable scene.
	For the BGK model, HL is a better choice to calculate the exact evolution equation.
	Meanwhile BK and TE scheme are more friendly to new collision models or whole collision approximation. What's more, BK and TE scheme have different convergent rates, in which BK scheme is faster.

	\begin{acknowledgments}
		The author would like to express his gratitude to Dr. Zecheng Gan for helpful discussion.
	\end{acknowledgments}

	\bibliography{EqvOfEM.bib}

\providecommand{\noopsort}[1]{}\providecommand{\singleletter}[1]{#1}%
\begin{thebibliography}{19}%
\makeatletter
\providecommand \@ifxundefined [1]{%
 \@ifx{#1\undefined}
}%
\providecommand \@ifnum [1]{%
 \ifnum #1\expandafter \@firstoftwo
 \else \expandafter \@secondoftwo
 \fi
}%
\providecommand \@ifx [1]{%
 \ifx #1\expandafter \@firstoftwo
 \else \expandafter \@secondoftwo
 \fi
}%
\providecommand \natexlab [1]{#1}%
\providecommand \enquote  [1]{``#1''}%
\providecommand \bibnamefont  [1]{#1}%
\providecommand \bibfnamefont [1]{#1}%
\providecommand \citenamefont [1]{#1}%
\providecommand \href@noop [0]{\@secondoftwo}%
\providecommand \href [0]{\begingroup \@sanitize@url \@href}%
\providecommand \@href[1]{\@@startlink{#1}\@@href}%
\providecommand \@@href[1]{\endgroup#1\@@endlink}%
\providecommand \@sanitize@url [0]{\catcode `\\12\catcode `\$12\catcode
  `\&12\catcode `\#12\catcode `\^12\catcode `\_12\catcode `\%12\relax}%
\providecommand \@@startlink[1]{}%
\providecommand \@@endlink[0]{}%
\providecommand \url  [0]{\begingroup\@sanitize@url \@url }%
\providecommand \@url [1]{\endgroup\@href {#1}{\urlprefix }}%
\providecommand \urlprefix  [0]{URL }%
\providecommand \Eprint [0]{\href }%
\providecommand \doibase [0]{http://dx.doi.org/}%
\providecommand \selectlanguage [0]{\@gobble}%
\providecommand \bibinfo  [0]{\@secondoftwo}%
\providecommand \bibfield  [0]{\@secondoftwo}%
\providecommand \translation [1]{[#1]}%
\providecommand \BibitemOpen [0]{}%
\providecommand \bibitemStop [0]{}%
\providecommand \bibitemNoStop [0]{.\EOS\space}%
\providecommand \EOS [0]{\spacefactor3000\relax}%
\providecommand \BibitemShut  [1]{\csname bibitem#1\endcsname}%
\let\auto@bib@innerbib\@empty
\bibitem [{\citenamefont {Liu}\ \emph {et~al.}(2006)\citenamefont {Liu},
  \citenamefont {Zou}, \citenamefont {Shi}, \citenamefont {Tian}, \citenamefont
  {Zhang},\ and\ \citenamefont {Zheng}}]{LiuZou-178}%
  \BibitemOpen
  \bibfield  {author} {\bibinfo {author} {\bibfnamefont {H.}~\bibnamefont
  {Liu}}, \bibinfo {author} {\bibfnamefont {C.}~\bibnamefont {Zou}}, \bibinfo
  {author} {\bibfnamefont {B.}~\bibnamefont {Shi}}, \bibinfo {author}
  {\bibfnamefont {Z.}~\bibnamefont {Tian}}, \bibinfo {author} {\bibfnamefont
  {L.}~\bibnamefont {Zhang}}, \ and\ \bibinfo {author} {\bibfnamefont
  {C.}~\bibnamefont {Zheng}},\ }\href@noop {} {\bibfield  {journal} {\bibinfo
  {journal} {Int. J. Heat Mass Tran.}\ }\textbf {\bibinfo {volume} {49}},\
  \bibinfo {pages} {4672} (\bibinfo {year} {2006})}\BibitemShut {NoStop}%
\bibitem [{\citenamefont {Baochang}\ and\ \citenamefont
  {Zhaoli}(2003)}]{BaochangZhaoli-179}%
  \BibitemOpen
  \bibfield  {author} {\bibinfo {author} {\bibfnamefont {S.}~\bibnamefont
  {Baochang}}\ and\ \bibinfo {author} {\bibfnamefont {G.}~\bibnamefont
  {Zhaoli}},\ }\href@noop {} {\bibfield  {journal} {\bibinfo  {journal} {Int.
  J. Mod. Phys. B}\ }\textbf {\bibinfo {volume} {17}},\ \bibinfo {pages} {173}
  (\bibinfo {year} {2003})}\BibitemShut {NoStop}%
\bibitem [{\citenamefont {Ladd}\ and\ \citenamefont
  {Verberg}(2001)}]{LaddVerberg-186}%
  \BibitemOpen
  \bibfield  {author} {\bibinfo {author} {\bibfnamefont {A.~J.~C.}\
  \bibnamefont {Ladd}}\ and\ \bibinfo {author} {\bibfnamefont {R.}~\bibnamefont
  {Verberg}},\ }\href@noop {} {\bibfield  {journal} {\bibinfo  {journal} {J.
  Stat. Phys.}\ }\textbf {\bibinfo {volume} {104}},\ \bibinfo {pages} {1191}
  (\bibinfo {year} {2001})}\BibitemShut {NoStop}%
\bibitem [{\citenamefont {Sheikholeslami}\ \emph {et~al.}(2014)\citenamefont
  {Sheikholeslami}, \citenamefont {Gorji-Bandpy},\ and\ \citenamefont
  {Ganji}}]{SheikholeslamiGorji-Bandpy-163}%
  \BibitemOpen
  \bibfield  {author} {\bibinfo {author} {\bibfnamefont {M.}~\bibnamefont
  {Sheikholeslami}}, \bibinfo {author} {\bibfnamefont {M.}~\bibnamefont
  {Gorji-Bandpy}}, \ and\ \bibinfo {author} {\bibfnamefont {D.~D.}\
  \bibnamefont {Ganji}},\ }\href@noop {} {\bibfield  {journal} {\bibinfo
  {journal} {Powder Technol.}\ }\textbf {\bibinfo {volume} {254}},\ \bibinfo
  {pages} {82} (\bibinfo {year} {2014})}\BibitemShut {NoStop}%
\bibitem [{\citenamefont {Liu}\ and\ \citenamefont
  {Cheng}(2013)}]{LiuCheng-181}%
  \BibitemOpen
  \bibfield  {author} {\bibinfo {author} {\bibfnamefont {X.}~\bibnamefont
  {Liu}}\ and\ \bibinfo {author} {\bibfnamefont {P.}~\bibnamefont {Cheng}},\
  }\href@noop {} {\bibfield  {journal} {\bibinfo  {journal} {Int. J. Heat Mass
  Tran.}\ }\textbf {\bibinfo {volume} {64}},\ \bibinfo {pages} {1041} (\bibinfo
  {year} {2013})}\BibitemShut {NoStop}%
\bibitem [{\citenamefont {Semma}\ \emph {et~al.}(2008)\citenamefont {Semma},
  \citenamefont {Ganaoui}, \citenamefont {Bennacer},\ and\ \citenamefont
  {Mohamad}}]{SemmaElGanaoui-240}%
  \BibitemOpen
  \bibfield  {author} {\bibinfo {author} {\bibfnamefont {E.}~\bibnamefont
  {Semma}}, \bibinfo {author} {\bibfnamefont {M.~E.}\ \bibnamefont {Ganaoui}},
  \bibinfo {author} {\bibfnamefont {R.}~\bibnamefont {Bennacer}}, \ and\
  \bibinfo {author} {\bibfnamefont {A.~A.}\ \bibnamefont {Mohamad}},\
  }\href@noop {} {\bibfield  {journal} {\bibinfo  {journal} {Int. J. Therm.
  Sci.}\ }\textbf {\bibinfo {volume} {47}},\ \bibinfo {pages} {201} (\bibinfo
  {year} {2008})}\BibitemShut {NoStop}%
\bibitem [{\citenamefont {Mountrakis}\ \emph {et~al.}(2015)\citenamefont
  {Mountrakis}, \citenamefont {Lorenz}, \citenamefont {Malaspinas},
  \citenamefont {Alowayyed}, \citenamefont {Chopard},\ and\ \citenamefont
  {Hoekstra}}]{MountrakisLorenz-483}%
  \BibitemOpen
  \bibfield  {author} {\bibinfo {author} {\bibfnamefont {L.}~\bibnamefont
  {Mountrakis}}, \bibinfo {author} {\bibfnamefont {E.}~\bibnamefont {Lorenz}},
  \bibinfo {author} {\bibfnamefont {O.}~\bibnamefont {Malaspinas}}, \bibinfo
  {author} {\bibfnamefont {S.}~\bibnamefont {Alowayyed}}, \bibinfo {author}
  {\bibfnamefont {B.}~\bibnamefont {Chopard}}, \ and\ \bibinfo {author}
  {\bibfnamefont {A.~G.}\ \bibnamefont {Hoekstra}},\ }\href@noop {} {\bibfield
  {journal} {\bibinfo  {journal} {J. Comput. Sci.-Neth}\ }\textbf {\bibinfo
  {volume} {9}},\ \bibinfo {pages} {45} (\bibinfo {year} {2015})}\BibitemShut
  {NoStop}%
\bibitem [{\citenamefont {Meng}\ and\ \citenamefont
  {Guo}(2016)}]{MengGuo-2046}%
  \BibitemOpen
  \bibfield  {author} {\bibinfo {author} {\bibfnamefont {X.}~\bibnamefont
  {Meng}}\ and\ \bibinfo {author} {\bibfnamefont {Z.}~\bibnamefont {Guo}},\
  }\href@noop {} {\bibfield  {journal} {\bibinfo  {journal} {Int. J. Heat Mass
  Tran.}\ }\textbf {\bibinfo {volume} {100}},\ \bibinfo {pages} {767} (\bibinfo
  {year} {2016})}\BibitemShut {NoStop}%
\bibitem [{\citenamefont {Liu}\ and\ \citenamefont {He}(2017)}]{LiuHe-2047}%
  \BibitemOpen
  \bibfield  {author} {\bibinfo {author} {\bibfnamefont {Q.}~\bibnamefont
  {Liu}}\ and\ \bibinfo {author} {\bibfnamefont {Y.}~\bibnamefont {He}},\
  }\href@noop {} {\bibfield  {journal} {\bibinfo  {journal} {Physica A:
  Statistical Mechanics and its Applications}\ }\textbf {\bibinfo {volume}
  {465}},\ \bibinfo {pages} {742} (\bibinfo {year} {2017})}\BibitemShut
  {NoStop}%
\bibitem [{\citenamefont {Bhatnagar}\ \emph {et~al.}(1954)\citenamefont
  {Bhatnagar}, \citenamefont {Gross},\ and\ \citenamefont
  {Krook}}]{BhatnagarGross-95}%
  \BibitemOpen
  \bibfield  {author} {\bibinfo {author} {\bibfnamefont {P.~L.}\ \bibnamefont
  {Bhatnagar}}, \bibinfo {author} {\bibfnamefont {E.~P.}\ \bibnamefont
  {Gross}}, \ and\ \bibinfo {author} {\bibfnamefont {M.}~\bibnamefont
  {Krook}},\ }\href@noop {} {\bibfield  {journal} {\bibinfo  {journal}
  {Physical review}\ }\textbf {\bibinfo {volume} {94}},\ \bibinfo {pages} {511}
  (\bibinfo {year} {1954})}\BibitemShut {NoStop}%
\bibitem [{\citenamefont {Boesch}\ and\ \citenamefont
  {Karlin}(2013)}]{BoeschKarlin-2075}%
  \BibitemOpen
  \bibfield  {author} {\bibinfo {author} {\bibfnamefont {F.}~\bibnamefont
  {Boesch}}\ and\ \bibinfo {author} {\bibfnamefont {I.~V.}\ \bibnamefont
  {Karlin}},\ }\href@noop {} {\bibfield  {journal} {\bibinfo  {journal} {Phys.
  Rev. Lett.}\ }\textbf {\bibinfo {volume} {111}} (\bibinfo {year}
  {2013})}\BibitemShut {NoStop}%
\bibitem [{\citenamefont {He}\ \emph {et~al.}(1998)\citenamefont {He},
  \citenamefont {Chen},\ and\ \citenamefont {Doolen}}]{HeChen-173}%
  \BibitemOpen
  \bibfield  {author} {\bibinfo {author} {\bibfnamefont {X.}~\bibnamefont
  {He}}, \bibinfo {author} {\bibfnamefont {S.}~\bibnamefont {Chen}}, \ and\
  \bibinfo {author} {\bibfnamefont {G.~D.}\ \bibnamefont {Doolen}},\
  }\href@noop {} {\bibfield  {journal} {\bibinfo  {journal} {J. Comput. Phys.}\
  }\textbf {\bibinfo {volume} {146}},\ \bibinfo {pages} {282} (\bibinfo {year}
  {1998})}\BibitemShut {NoStop}%
\bibitem [{\citenamefont {He}\ and\ \citenamefont {Luo}(1997)}]{HeLuo-2016}%
  \BibitemOpen
  \bibfield  {author} {\bibinfo {author} {\bibfnamefont {X.~Y.}\ \bibnamefont
  {He}}\ and\ \bibinfo {author} {\bibfnamefont {L.~S.}\ \bibnamefont {Luo}},\
  }\href@noop {} {\bibfield  {journal} {\bibinfo  {journal} {Phys. Rev. E}\
  }\textbf {\bibinfo {volume} {56}},\ \bibinfo {pages} {6811} (\bibinfo {year}
  {1997})}\BibitemShut {NoStop}%
\bibitem [{\citenamefont {Ye}\ \emph {et~al.}(2017)\citenamefont {Ye},
  \citenamefont {Kuang},\ and\ \citenamefont {Yang}}]{YeKuang-2092}%
  \BibitemOpen
  \bibfield  {author} {\bibinfo {author} {\bibfnamefont {H.}~\bibnamefont
  {Ye}}, \bibinfo {author} {\bibfnamefont {B.}~\bibnamefont {Kuang}}, \ and\
  \bibinfo {author} {\bibfnamefont {Y.}~\bibnamefont {Yang}},\ }\href@noop {}
  {\enquote {\bibinfo {title} {Derivation of the lattice boltzmann equation via
  method of characteristics},}\ } (\bibinfo {year} {2017}),\ \Eprint
  {http://arxiv.org/abs/1703.05889} {arXiv:1703.05889 [physics.flu-dyn]}
  \BibitemShut {NoStop}%
\bibitem [{\citenamefont {Shan}(2016)}]{Shan-2088}%
  \BibitemOpen
  \bibfield  {author} {\bibinfo {author} {\bibfnamefont {X.}~\bibnamefont
  {Shan}},\ }\href@noop {} {\bibfield  {journal} {\bibinfo  {journal} {J.
  Comput. Sci.-Neth}\ }\textbf {\bibinfo {volume} {17}},\ \bibinfo {pages}
  {475} (\bibinfo {year} {2016})}\BibitemShut {NoStop}%
\bibitem [{\citenamefont {Hwang}(2016)}]{Hwang-2023}%
  \BibitemOpen
  \bibfield  {author} {\bibinfo {author} {\bibfnamefont {Y.}~\bibnamefont
  {Hwang}},\ }\href@noop {} {\bibfield  {journal} {\bibinfo  {journal} {J.
  Comput. Phys.}\ }\textbf {\bibinfo {volume} {322}},\ \bibinfo {pages} {52}
  (\bibinfo {year} {2016})}\BibitemShut {NoStop}%
\bibitem [{\citenamefont {Yong}\ \emph {et~al.}(2016)\citenamefont {Yong},
  \citenamefont {Zhao},\ and\ \citenamefont {Luo}}]{YongZhao-2019}%
  \BibitemOpen
  \bibfield  {author} {\bibinfo {author} {\bibfnamefont {W.~A.}\ \bibnamefont
  {Yong}}, \bibinfo {author} {\bibfnamefont {W.}~\bibnamefont {Zhao}}, \ and\
  \bibinfo {author} {\bibfnamefont {L.~S.}\ \bibnamefont {Luo}},\ }\href@noop
  {} {\bibfield  {journal} {\bibinfo  {journal} {Phys. Rev. E}\ }\textbf
  {\bibinfo {volume} {93}},\ \bibinfo {pages} {33310} (\bibinfo {year}
  {2016})}\BibitemShut {NoStop}%
\bibitem [{\citenamefont {Dellar}(2013)}]{Dellar-2074}%
  \BibitemOpen
  \bibfield  {author} {\bibinfo {author} {\bibfnamefont {P.~J.}\ \bibnamefont
  {Dellar}},\ }\href@noop {} {\bibfield  {journal} {\bibinfo  {journal}
  {Comput. Math. Appl.}\ }\textbf {\bibinfo {volume} {65}},\ \bibinfo {pages}
  {129} (\bibinfo {year} {2013})}\BibitemShut {NoStop}%
\bibitem [{\citenamefont {Arfken}\ \emph {et~al.}(2011)\citenamefont {Arfken},
  \citenamefont {Weber},\ and\ \citenamefont {Harris}}]{ArfkenWeber-2095}%
  \BibitemOpen
  \bibfield  {author} {\bibinfo {author} {\bibfnamefont {G.~B.}\ \bibnamefont
  {Arfken}}, \bibinfo {author} {\bibfnamefont {H.~J.}\ \bibnamefont {Weber}}, \
  and\ \bibinfo {author} {\bibfnamefont {F.~E.}\ \bibnamefont {Harris}},\
  }\href@noop {} {\emph {\bibinfo {title} {Mathematical Methods for Physicists:
  A Comprehensive Guide}}}\ (\bibinfo  {publisher} {Academic Press},\ \bibinfo
  {year} {2011})\BibitemShut {NoStop}%
\end{thebibliography}%
	
	\end{document}